\documentclass[aps,pra,twocolumn,amsmath,amssymb,superscriptaddress,showpacs]{revtex4-1}

\usepackage{amsfonts}
\usepackage{amsmath}
\usepackage{amssymb}
\usepackage{amsthm}
\usepackage{fontenc}
\usepackage{graphicx}
\usepackage{xcolor}
\usepackage{textcomp}
\usepackage{epstopdf}
\usepackage{braket}
\usepackage{mathtools}
\usepackage{amsmath}
\usepackage{dcolumn}
\usepackage{multirow}

\begin{document}
\title{  
All-electrical production of spin-polarized currents in carbon nanotubes: Rashba spin-orbit interaction}

\author{Hern\'an Santos}
\affiliation{Departamento de F\'{\i}sica Fundamental, Universidad Nacional de Educaci\' on a Distancia, Apartado 60141, E-28040 Madrid, Spain}
\author{A. Latg\' e}
\affiliation{Instituto de F\' isica, Universidade Federal Fluminense, Niter\' oi, Av.~Litor\^anea sn 24210-340, RJ-Brazil}
\author{J. E. Alvarellos}
\affiliation{Departamento de F\'{\i}sica Fundamental, Universidad Nacional de Educaci\' on a Distancia, Apartado 60141, E-28040 Madrid, Spain}
\author{Leonor Chico}
%\email[A. Latg\'e]{andrea.latge@gmail.com}
\affiliation{Instituto de Ciencia de Materiales de Madrid, Consejo Superior de Investigaciones Cient\'{\i}ficas, C/ Sor Juana In\'es de la Cruz 3, 28049 Madrid, Spain}

\date{\today}

\begin{abstract}

We study the effect of the Rashba spin-orbit interaction in the quantum transport 
of carbon nanotubes with arbitrary chiralities. 
For certain spin directions, we find a strong spin-polarized electrical current 
that depends on the diameter of the tube, the length of the Rashba region and 
on the tube chirality. 
Predictions for the spin-dependent conductances are presented for different families 
of achiral and chiral tubes. We have found that different symmetries acting on spatial and 
spin variables have to be considered in order to explain the relations between spin-resolved conductances in carbon nanotubes. 
These symmetries are more general than those employed in planar graphene systems. 
Our results indicate the possibility of having stable spin-polarized electrical currents in 
absence of external magnetic fields or magnetic impurities in carbon nanotubes. 
 
\end{abstract}
%\pacs{ 73.63.-b, 72.25-b}

\maketitle

\section{Introduction}
Spintronics relies on a sensitive manipulation of the interaction between the particle spin and its environment to produce spin currents \cite{SPINTRONICS04}. At the nanoscale,  the production of strong polarized currents 
is being explored extensively in different systems \cite{Crook2006,Debray2009,Tang2015}. The most immediate approach is the use of external magnetic fields, but they are hard to control at the nanoscale. A promising way is to employ materials and devices capable of producing 
  spin-polarized electrical currents without external magnetic fields, by means of strong spin-orbit interactions (SOI), such as two-dimensional transition-metal dichalcogenides 
  \cite{Wang2012,Yuan2014} or semiconductors \cite{Debray2009,Yin2013,Chuang2015}. 
  Novel analogues of graphene, such as silicene, germanene and phosphorene, share many of its interesting properties, but with more promising features in this respect \cite{Vogt2012,Tsai2013,Balendhran2015}. 
 These materials have been additionally proposed for their use in valleytronics, as well as for quantum-confined structures exhibiting interesting spin-dependent transport properties \cite{Loss2013}. Indeed, the exploitation of spin-orbit effects for electronic applications has resulted in the greatly active research field of spin-orbitronics \cite{Manchon2015}. 

Graphene, a material that presents a wealth 
 of groundbreaking applications, was seminally proposed in this spintronics scenario as a possible quantum spin Hall insulator \cite{Kane2005}. 
However, due to its small intrinsic SOI, it is unlikely to observe the quantum spin Hall effect in graphene edges. Rather, it presents characteristic long spin-diffusion lengths observed at room temperature \cite{Han2010}.  In fact, active routes to modify graphene by hybridization \cite{Dedkov2008,Marchenko2012,Gmitra2013}  or by strain \cite{Levy2010,Lim2015} exist.
In general, carbon nanostructures show an enhanced 
SOI effect when the atoms are arranged in a cylindrical fashion such as in carbon nanotubes (CNTs). Since the pioneering work of Ando \cite{Ando2000}, several theoretical works have explored the role of spin-orbit interactions in the electronic structure of carbon nanotubes and curved carbon systems, such as bent and folded graphene ribbons, in which the effects are enhanced because of curvature  \cite{Chico2004,Huertas2006,Chico2009,ISS09,HSantos1,Costa2013}.  Due to this increase, spin-orbit effects could be measured in carbon nanotubes  \cite{KIRM08,STEELE13,Laird2015}, opening the way for their application in quantum computation \cite{Flensberg2010,Laird2013} and spintronic devices \cite{Kuemmeth2010}. 

The Rashba spin-orbit (RSO) effect is expected when a structural inversion asymmetry is produced \cite{BycRashba84,SPINTRONICS04,WinklerBook}. There are several ways to achieve such asymmetry in carbon nanostructures. 
A successful route is to consider heavy adatoms randomly spread on a graphene lattice \cite{Weeks2011}. In this line, spin angle-resolved photoemission spectroscopy
experiments suggest that a large Rashba-type SOI can be tuned in graphene over metals such as Au or Ni 
% by the application of an external electric field % the electric field was not external, it was effective (built-in)
 \cite{Dedkov2008,Varykhalov2008}. 
Likewise, the presence of adatoms \cite{Xu2007} or adsorbed molecules in the surface of CNTs, such as as DNA strands \cite{Diniz2012}, enhances the Rashba interaction due to 
the combined effect of rehybridization and breaking inversion asymmetry. Recently, it has been experimentally verified that DNA-wrapped CNTs act as spin filters due to 
Rashba spin-orbit interaction \cite{Alam2015}.

%\begin{figure}[b]
\begin{figure}[h]
\includegraphics[width=1.00\columnwidth]{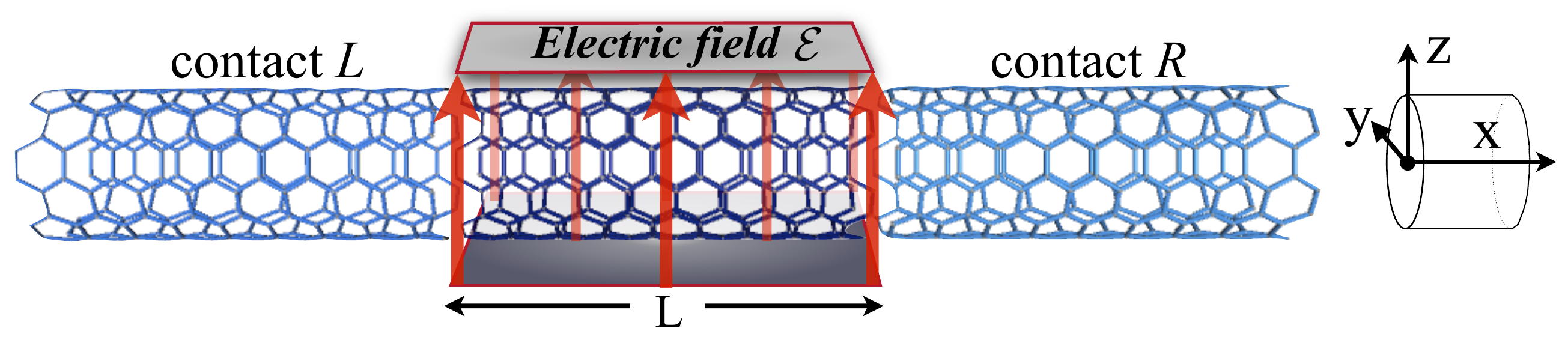}
\caption{(Color online) 
Schematic view of the device geometry. 
Left (L) and right (R) contacts are pristine CNTs without RSO interaction. 
The central part of the device is a conducting CNT of length $L$, with Rashba 
spin-orbit interaction induced by the presence of an electric field in 
the $z$ direction (red arrows) generated by two capacitor layers.  
The directions $x$, ${y}$ and ${z}$ are defined in the right part of the figure.
}
\label{device}
\end{figure}

In pure graphene, intrinsic spin-orbit effects are negligible \cite{Min2007}. As mentioned above, curvature effects increase spin-orbit splittings in carbon nanotubes to the meV range, allowing their measurement \cite{KIRM08,STEELE13}. Other strategies, such as hydrogenation, proximity effects or chemical functionalization, have been 
experimentally shown to boost the values of spin-orbit interaction in graphene \cite{Marchenko2012,Balakrishnan2014,Calleja2015}. For instance, in graphene grown by chemical vapor deposition spin-orbit splittings around 20 meV were measured, due to the presence of Cu atoms \cite{Balakrishnan2014}. Intercalation of Au in graphene grown on Ni has been reported to produce splittings up to 100 meV due to hybridization with Au atoms \cite{Marchenko2012}, and even larger values have been attributed to the presence of Pb in graphene samples \cite{Calleja2015}. It is reasonable to expect that such hybridization and proximity effects can enhance the strength of spin-orbit interaction in carbon nanotubes, rendering values of interest for spintronics. 

However, the increase of SOI effects in order to allow all-electrical control of spin currents conveys an important drawback: it also affects the spin lifetime. One possible strategy to correct this problem
% : either to equilibrate Rashba SOI with a Dresselhaus term, an approach not always possible for all material, or 
is to lower the dimension of these systems, given that in one-dimensional structures spin scattering is reduced. 
For example, semiconductor quantum wires have been proved to present an enhanced and tunable Rashba effect in different experimental setups \cite{Liang2012}. In this regard, CNTs emerge as interesting candidates that, similarly to quantum wires, have a non-planar geometry. 
Therefore, a plausible alternative to get structural inversion asymmetry and the subsequent 
Rashba splitting in carbon nanotubes is to apply an electric field perpendicular to the tube axis \cite{Martino2005,Klinovaja2011}. 

If a spin-polarized current from a ferromagnetic electrode is injected in the nanotube, it will operate as a Datta-Das transistor \cite{Datta1990}:  
depending on the strength of the applied electric field, the spin direction of the output current can be changed \cite{Martino2005}. 
Another possibility is to exploit the Rashba interaction to obtain a spin filter, i.e., a spin-polarized conductance from an unpolarized input current without the use of magnetic fields or ferromagnetic electrodes, by all-electrical means.

We explore this route for the obtention of spin-polarized currents in carbon nanotubes. In a previous work, we have studied the role of symmetries in the spin-resolved conductance of
planar systems, particularizing for graphene nanoribbons \cite{Chico2015}. 
Here we extend our study to the non-planar 
geometries of carbon nanotubes, analyzing different tube chiralities 
and extending our symmetry analysis to the corresponding three-dimensional geometries. 
Our main results are the following:

\noindent
(i) We show that for the optimal configuration which maximizes the spin polarization of the conductance, namely, with the electric field, the current and the spin projection direction in perpendicular directions, all CNTs present a remarkable spin-polarized current irrespectively of chirality. This effect is robust with respect to the length of the Rashba region and the radius of the nanotube. 

\noindent
(ii) We have checked that the one-orbital tight-binding approximation gives a very good description of the spin-resolved conductances, with minor numerical differences with respect to the four-orbital tight-binding model.   

\noindent
(iii) We find that more general symmetries that those applied in graphene nanoribbons are needed to account for the relations between the spin-resolved conductances in carbon nanotubes. In the case of planar systems, it was sufficient to consider symmetry operations acting simultaneously in spatial and spin spaces, whereas for carbon nanotubes we find that the relevant symmetries may act in spatial and spin variables independently. 

\noindent
(iv) Besides suggesting a possible path for spin devices without magnetic fields based in carbon nanotubes, our symmetry reasoning might be of use for other systems with Rashba spin-orbit interaction,  in which these symmetries can play a role.

This work is organized as follows. In Sec. \ref{sec:model} we describe the geometry of the system, the model employed for the carbon nanotubes and the computation of the spin transport properties. Sec. \ref{sec:spt} presents the results for spin-polarized conductances in the optimal geometry, with the electric field, the current and the spin projection direction in perpendicular directions. Sec. \ref{sec:sym} details the symmetry relations for the spin-resolved conductances in carbon nanotubes with a Rashba term for all spin polarization directions. In Sec. \ref{sec:sum} we summarize our results and draw our conclusions. Finally, the Appendix presents the spatial symmetries of finite-sized carbon nanotubes. 

\section{System and model}
\label{sec:model}
\subsection{Description of the system}

The proposed system to obtain spin-polarized currents is composed of a pristine carbon nanotube with a finite central region of length $L$ placed between two metallic 
gates which can apply an external electric field, which produces a tunable RSO interaction 
in this part of the nanotube.   
A schematic view of the CNT device is shown in Fig. \ref{device}. 
The electric field $\boldsymbol{\mathcal{E}} $ inducing the Rashba spin-orbit interaction is taken in the direction perpendicular to the tube axis ($z$-direction), and it is 
indicated 
by red arrows.
Thus, the Rashba region corresponds to the portion of the tube under 
the effect of the electric field.

We study several CNTs of different chiralities, radii and length of the Rashba 
region, considering different spin polarization directions ($x$, $y$, and ${z}$). 
As we are interested in transport properties, we concentrate in metallic 
nanotubes, with nanotube indices $(n,m)$ verifying $n-m=3q$, 
with $q$ integer including 0. 

For an $(n,m)$ CNT, its radius $R$ is given by  
$R=\frac{\sqrt{3}a_{c}}{2\pi}(n^2+m^2+nm)^{1/2}$, 
where $a_{c}$ is the carbon bond length in graphene. 
We give the length of the Rashba region in translational unit cells, so that $L = NT$, 
with $T=3a_c$ and $\sqrt 3 a_c$ for zigzag and armchair tubes, respectively. 
For chiral nanotubes, $L$ can be substantially larger \cite{Lchiral}. 

\subsection{Model and method}

Most of our calculations are performed within the one-orbital tight-binding model employed in Ref. \onlinecite{Chico2015}. 
%however, as in carbon nanotubes curvature has a relevant contribution, we will compare 
%this simple model to a four-orbital hamiltonian taking into account curvature effects. 
The Hamiltonian of an undoped CNT with Rashba spin-orbit coupling in the nearest-neighbor hopping tight-binding approximation \cite{Qiao2010,Lenz2013} can be written as $H=H_0 + H_R$, 
where  $H_0$ is the kinetic energy term,  
$H_0 = -\gamma_0 \sum
%_{\substack{<i,j>\\\alpha}}
 c_{i\alpha}^\dagger  c_{j\alpha}$, 
%\end{equation}
with $\gamma_0$ being the nearest-neighbor hopping and $c_{i\alpha}$,  $c_{j\alpha}^\dagger$  the destruction and creation operators for an electron with spin projection $\alpha$ in site $i$ and $j$, respectively. The 
%Rashba SOI 
RSO 
contribution is given by
\begin{equation} 
H_R= %\gamma_0 \sum_{\substack{<i,j>\\\alpha}} c_{i\alpha}^\dagger  c_{j\alpha} + 
\frac{i \lambda_R}{a_{c}}\sum_{\substack {<i,j>\\\alpha,\beta}} c_{i\alpha}^\dagger  
\big[ (\boldsymbol{\sigma} \times \bold{d}_{ij}) \cdot \bold {e}_z \big]_{\alpha \beta} 
\ c_{j\beta} \,\, ,
\label{HR} 
\end{equation}
with  $\boldsymbol\sigma$ being the Pauli spin matrices, $\bold{d}_{ij}$ the position vector between sites ${i}$ and ${j}$, and $\alpha, \beta$ are the spin projection indices. 
The electric field is along the unit vector $\bold {e}_z$.
The Rashba spin-orbit strength  $\lambda_R$ is given by the electric field intensity, 
its sign defined by the sense of the field.
A value $\lambda_R = 0.1 \gamma_0$ will we used in this work, the same as in Ref. \onlinecite{Chico2015}. 
Note that, although this value is very large with respect to the 
SOI of a pristine carbon nanotube, the enhancements reported for functionalized or decorated graphene 
\cite{Marchenko2012,Balakrishnan2014,Calleja2015} discussed above indicate that similar values may be achieved in 
carbon nanotubes with analogous techniques.

The conductance ${G}^{LR}_{\sigma \sigma'}(E)$ is calculated in the Landauer approach, 
in the zero bias approximation, 
where it is proportional to the probability that one 
electron with spin $\sigma$ and energy $E$ in electrode $L$ reaches electrode $R$ with spin  
$\sigma'$,
by using the Green function formalism \cite{Xu2007,Diniz2012}. 
The spin-resolved conductance along the nanotube axis ($x$ direction) is then given by
\begin{equation}
{G}^{LR}_{\sigma \sigma'}(E) = \frac{e^2}{h} \ 
\mathrm{Tr} \big[ \Gamma^{L}_{\sigma}g^r_{\sigma \sigma^\prime}\Gamma^{R}_{\sigma'}g^a_{\sigma' \sigma} \big] \,\,. 
\label{GLR}
\end{equation}
Here  $g^{r(a)}_{\sigma \sigma^\prime} $  is the retarded (advanced) Green 
function of the conductor and 
$\Gamma^{L(R)}_{\sigma}=i \big[ \sum^{r}_{L(R) \sigma} - \sum^{a}_{L(R) \sigma} \big]$ 
is written in terms of the left (right) lead selfenergies 
$\Sigma^{r(a)}_{L(R) \sigma}$.

Assuming left to right conduction, we define the spin polarization of the conductance
in the $i$-direction  ($i = x, y,$ and $z$)  as 
\begin{equation} 
P_{i}(E) = G^{LR}_{\sigma_i \sigma_i} - G^{LR}_{\sigma_i  \bar \sigma_i} + 
G^{LR}_{\bar \sigma_i \sigma_i} - G^{LR}_{\bar \sigma_i \bar \sigma_i}\, , 
\end{equation}
where $\bar \sigma_i$ stands for  $-\sigma_i$. Note that  $i$ is the direction into which the spin is projected. In the small-bias limit, this magnitude is proportional to the $i$-component of the spin current. Other authors employ a normalized adimensional polarization factor \cite{Zhai2005,Diniz2012}; we choose this definition which is directly related to the most commonly measured transport quantity in low-dimensional systems, the conductance. 

% is this last sentence superfluous? is thus the direction into which the spin is projected. 

We suppose that an unpolarized charge current flows from the left electrode 
through the Rashba region towards the right electrode. 
In the Rashba region, spin scattering takes place, giving rise to a 
spin-polarized current {\it for certain spin projection directions and system symmetries}. 
Note that due to the analytical expression for $H_R$, if  $\lambda_R$ changes 
sign (i. e., when inverting the sense of the electric field) 
the spin polarizations on the $x$ and $y$ directions also change their sign, 
whereas the polarization in the perpendicular direction ($z$ axis) is unaffected.

We will also use a four-orbital Hamiltonian able to take into account the effects 
of curvature, in order to verify whether it has 
any relevant contribution when compared to the simpler one-orbital model.

\section{Spin-polarized currents in carbon nanotubes}
\label{sec:spt}

For planar devices composed of graphene nanoribbons, and just from symmetry considerations it is possible to conclude in which spin direction 
a non-zero spin-polarized current can be obtained \cite{Chico2015}. 
Moreover, it can be also deduced that the optimal spin projection direction to achieve  the largest 
spin-polarized current is perpendicular to both the current direction and the external 
electric field,
as it can be inferred from the Rashba term written in the continuum approximation, 
$H_R \propto ({\boldsymbol \sigma}  \times {\bold k}) \cdot \boldsymbol{\mathcal{E}} $, where ${\bold k}$ is the wavevector of the carrier \cite{WinklerBook}.

So, we start by studying the spin projection direction for which the 
spin-polarized current is expected to be largest: 
%perpendicular to both the electric field and current, which is 
$y$ for our choice of axes, as 
shown in Fig. \ref{device}.  
For this spin projection direction, we always obtain a nonzero polarization 
irrespectively of the geometry of the nanotube, 
as it also happens for planar systems \cite{Chico2015}.
%Similar values for the radii and RSO lengths are considered.

\begin{figure}[h]
\includegraphics[width=\columnwidth]{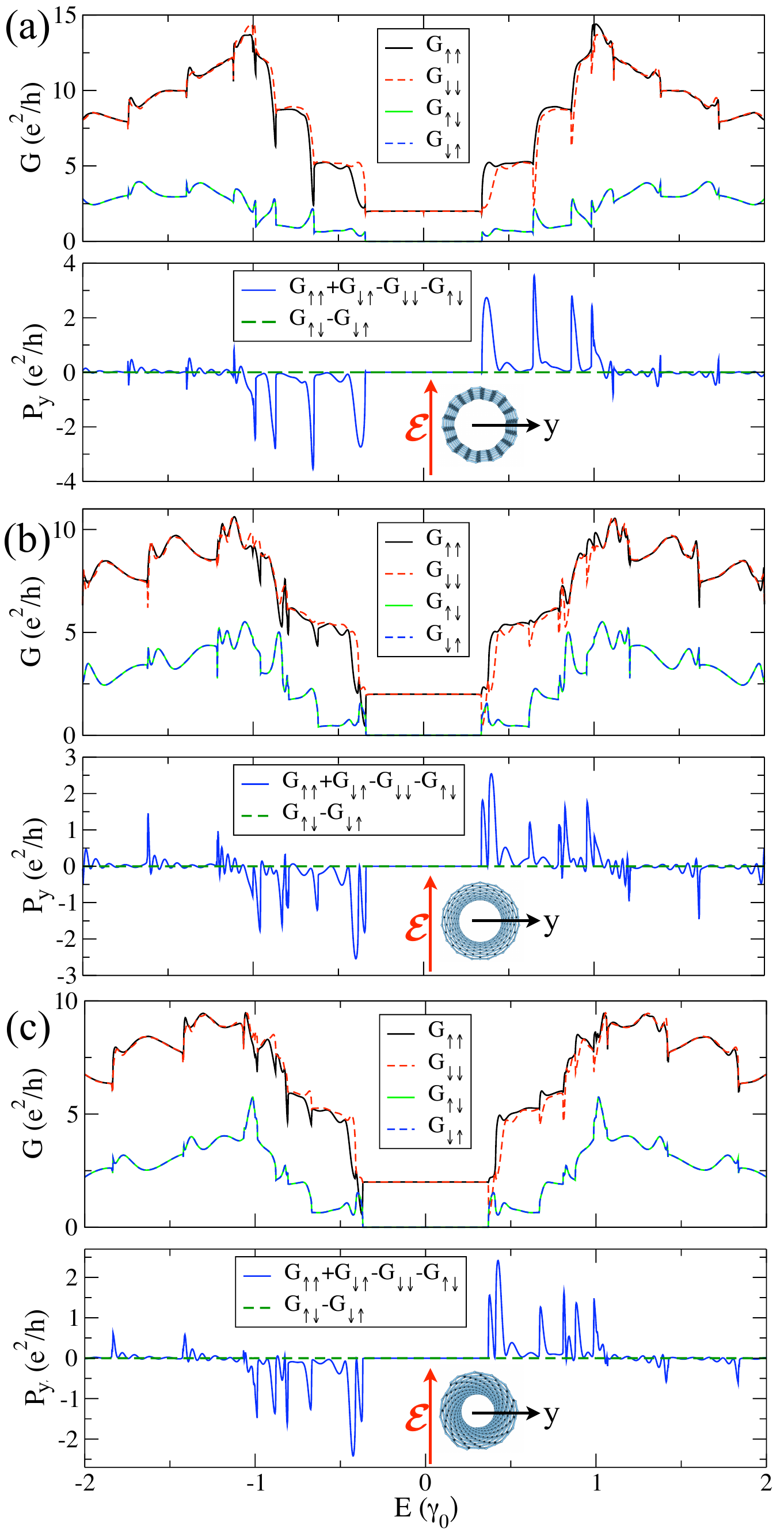}
\caption{(Color online) 
Spin-dependent conductances (top panels) and  spin polarization of the conductances as functions of the carrier energy 
(bottom panels) of  
(a) armchair (9,9) CNT with $N=16$ central unit cells under RSO interaction; 
(b) zigzag (15,0) CNT with $N= 9$; 
and (c) chiral (12,3) CNT with $N=6$. 
The electric field ($z$) and spin polarization ($y$) 
directions are shown in the insets.
}
\label{fig2}
\end{figure}

\subsection{ Rashba spin-orbit interaction and chirality}

Fig. \ref{fig2} shows the spin-dependent conductance and spin polarization as functions of the energy of the carriers of three instances of carbon nanotubes with different chiralities but similar radii and  
Rashba region lengths:
an armchair (9,9) and a zigzag (15,0) achiral CNTs, 
and a chiral tube (12,3).
It is noteworthy that for all tubes the polarization is null for energies close to zero. 
This is because for this range of energies there is only one open channel in the 
output (right) lead,
so the SOI does not produce any polarization due to symmetry and current conservation reasons \cite{Zhai2005}.
When the second valence or conduction band of the nanotube leads enters, a sharp spin polarization of the conductance is produced. 
The details of the positions and the number of maxima of the spin polarizations 
%currents 
can be related to the onset of the specific conductance channels, 
and therefore to the CNT chirality, as it can be inferred from the spin-resolved 
conductance panels. 
The polarization results show electron-hole symmetry; $P_{y}(E)=-P_{y}(-E)$. 

Note that for this spin projection direction, the sign of the polarization at a 
fixed electronic energy can be changed by flipping the electric field, 
because the analytical expression for $H_R$ in (\ref{HR}) 
makes the spin polarization 
on the $y$ direction to change sign,
suggesting another way for the application of Rashba CNT systems.

%\begin{table}[ht]
\begin{table}[b]
\renewcommand{\arraystretch}{2}
\setlength{\tabcolsep}{6pt} 
\caption{Nanotube indices, chiral angle, radius $(R)$, and length of the Rashba region $(L)$ for the set of nanotubes presented in Fig. \ref{fig3}.}
\begin{center}
\begin{tabular}{|c|c|c|c|}
%\begin{tabular}{cccc}
%\begin{tabular}{ccc}
      \hline
%  CNT indices  &  Chiral angle   &  Radius  & Rashba length \\
  $(n,m)$  &  $\theta(^{\rm o})$  & $ R$(\AA) & $L$(\AA)  \\
   \hline
     \hline
      (9,9) & 30 & 6.10 & 39.36 \\
     \hline 
     (8,8) & 30 & 5.42 & 39.36 \\ 
      \hline
     (10,7) & 24.2 & 5.79 & 42.04 \\
      \hline 
     (11,5) & 17.8 & 5.55 & 40.28 \\
      \hline
      (12,3) & 10.9 & 5.38 & 39.06 \\
     \hline
     (13,1) & 3.7 & 5.30 & 38.40  \\
     \hline
      (15,0) &0 & 5.87 & 38.34 \\
\hline
\end{tabular}
\end{center}
\label{tableI}
\end{table}

All the polarization of the conductances presented in Fig. \ref{fig2} mainly stem from 
the difference between 
$G_{\uparrow\uparrow}$ and 
$G_{\downarrow\downarrow}$ conductances,
but the difference $G_{\uparrow \downarrow} - G_{\downarrow \uparrow}$ 
is exactly zero in the (9,9) and (15,0) cases whereas is negligible for the (12,3) CNT.
As it is known from previous works \cite{Zhai2005,Chico2015}, 
these results should be ultimately related to the symmetries of the systems, 
but we postpone the discussion for the next Section.

\begin{figure}[t]
\includegraphics[width=0.95\columnwidth]{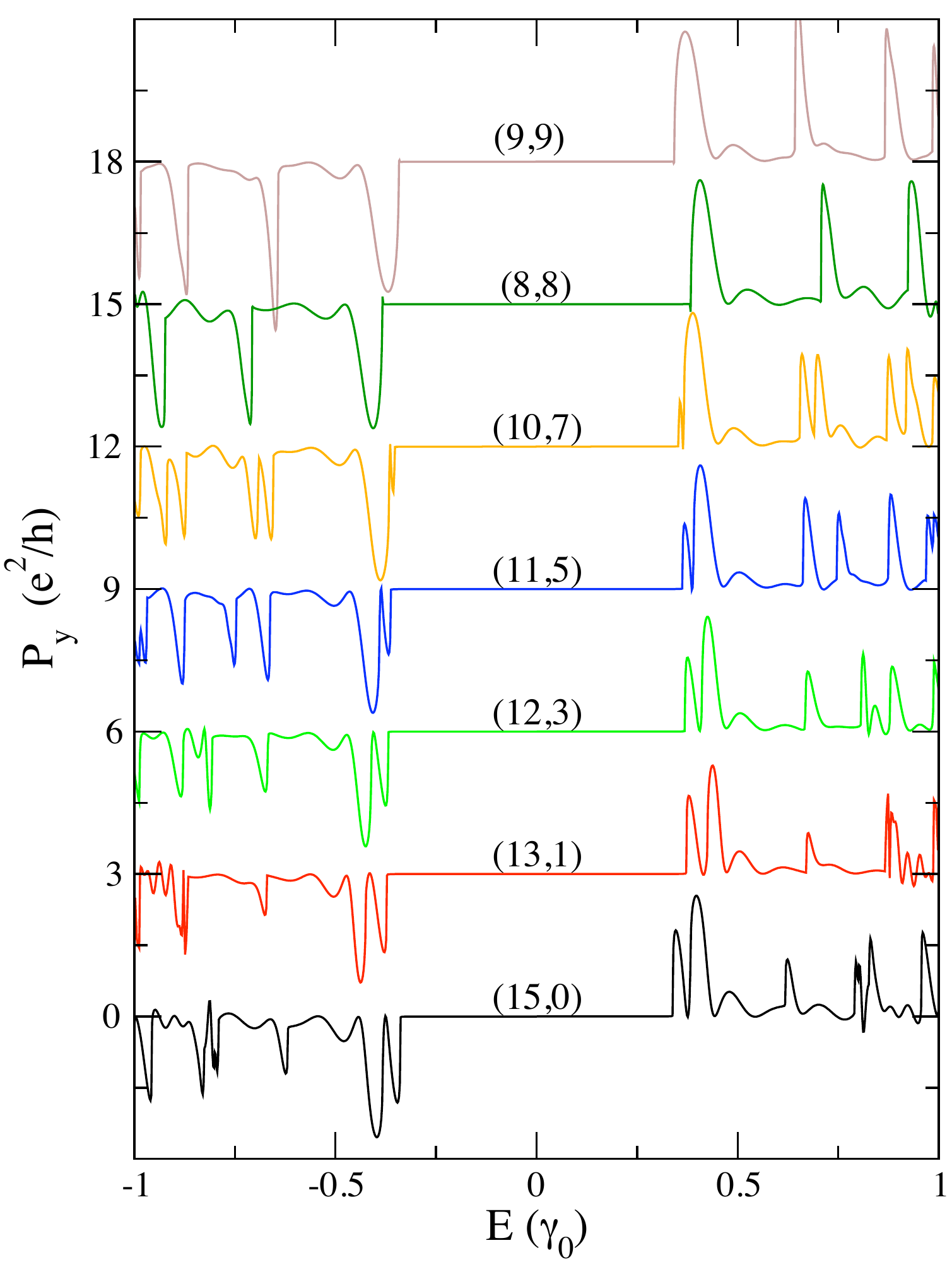}
\caption{(Color online) 
Spin-dependent polarization currents (for spins projected in the ${y}$-direction) 
for several carbon nanotubes with different chiralities but with 
comparable radii and very similar lengths $L$, as given in Table \ref{tableI}. }
\label{fig3}
\end{figure}

\begin{figure}[t]
\includegraphics[width=\columnwidth]{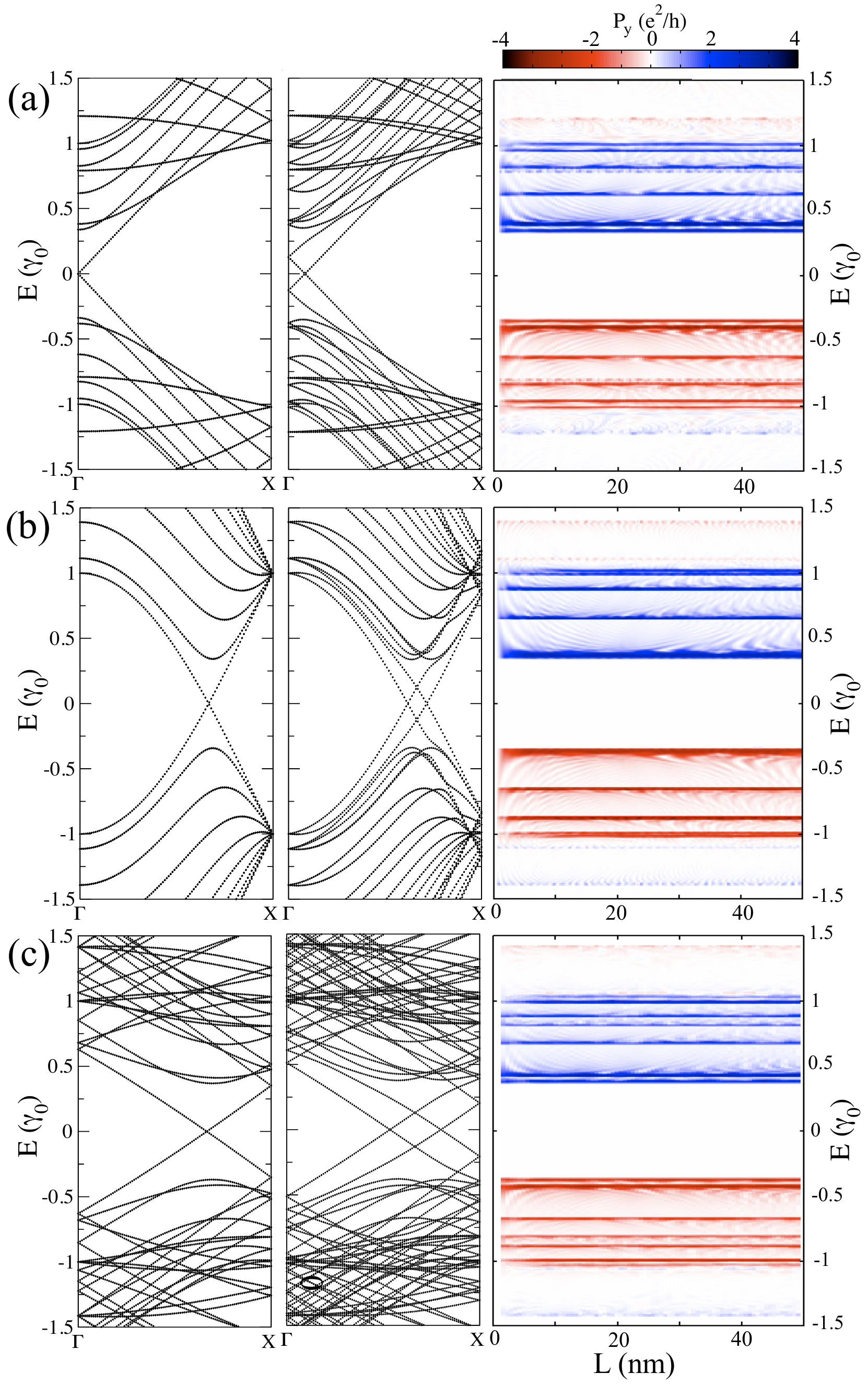}
\caption{(Color online) 
Electronic band structures of (a) (15,0), (b) (9,9) and (c) (12,3) CNTs without RSO interaction effects 
(left panels) and with RSO (center panels). 
The right panels show spin-dependent polarization maps,  $P_y (E,L)$, as functions of 
the energy and the Rashba region length. The spin polarization intensity 
in units of $e^2/h$ is plotted in a color scale. % (ranging from -4 to 4).
}
\label{fig4}
\end{figure}

Fig.  \ref{fig2} also shows that the maximum values of the spin conductance polarizations %currents 
are similar for the three nanotubes chosen, 
a hint that chirality does not play a crucial role 
{\em for this spin projection direction} in this respect.
In order to check this observation more carefully, we present in Fig. \ref{fig3} 
the spin polarization of several tubes with various chiralities but 
with similar radii and length of the Rashba region.
The geometric details of the tubes considered are presented in Table I.  
The values of the maxima and their energy positions are 
quite close for all the CNTs considered. 
%those maximun values of the current polarization 
%closer to the Fermi level, and their energy positions, 
%are very close for all the CNTs considered. 
The most important difference for the bumps closer to the Fermi level of the pristine tube (zero energy) 
is that they evolve into two split maxima when the chirality of the tubes varies 
from $\theta=30^{\rm o}$ [(8,8) CNT] 
to $\theta=0^{\rm o}$ [(15,0) CNT]. 
This splitting is related to the appearance of new channels in the pristine 
conductance of the corresponding tubes, which obviously depend on the chirality. 
In all the results we have found that the beginning of each bump coincides 
with the entrance of a new electronic band of the pristine nanotube constituting 
the leads. 
This fact is illustrated in Fig. \ref{fig4}, that presents the band structures of three 
tubes with different chiralities (the corresponding radii are given in Table \ref{tableI}).
The spin polarization of the conductances 
are also drawn in the right panels as a color map, $P_y (E,L)$. 
These panels clearly show that $P_y (E,L)$ is much greater when the energy 
coincides with a new band of the pristine system, that is, when a 
new conductance channel opens. 
The second and third band above (or below) the Fermi level of both, zigzag 
and chiral CNTs, are very close in energy, giving rise to a double peak in the 
polarization. 
The maximum values are around $P_y  \sim 4$ $e^2/h$.  

For fixed energies between the maxima of $P_y (E)$, 
Fig. \ref{fig4} shows an oscillating behavior of the polarization in terms 
of the length of the Rashba region, $P_y (L)$.
These oscillations are related to the coupling between the electronic states 
inside the RSO region and the electronic states of the leads, where no 
SOI is considered.  
For all the CNTs considered, %As these oscillations could \texttt{greatly} vary the values of $P_y$, 
the spin scattering Rashba region could behave like a spin polarizer, specially for the energies of the polarization bumps.  
In these energy intervals, it can be possible to obtain a strongly polarized output current from a 
non-polarized input current in a pristine CNT.

\begin{figure}[t]
\includegraphics[width=\columnwidth]{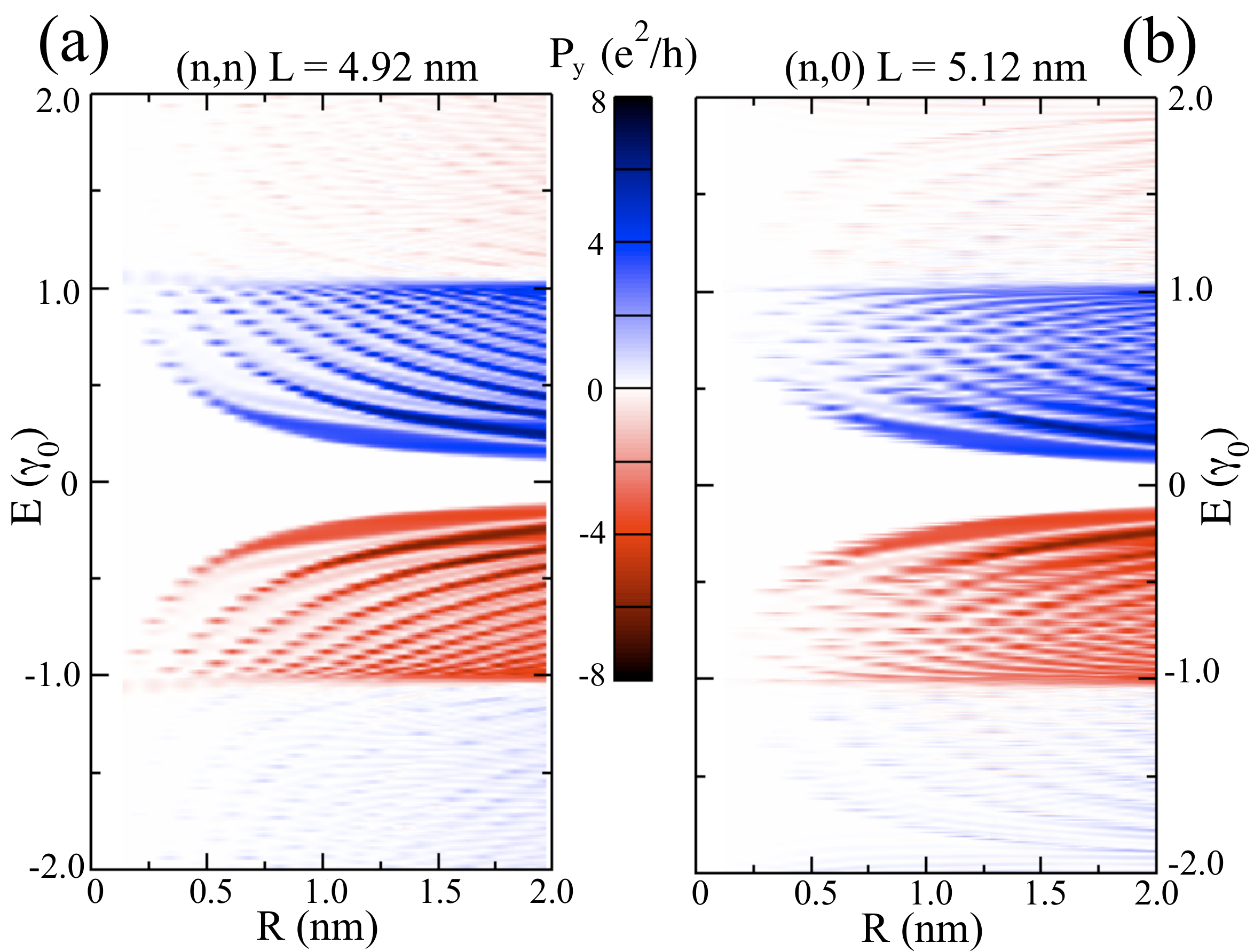}
\caption{(Color online) 
Spin-dependent polarization maps,  $P_y (E,R)$, as functions of the 
energy and the tube radius for achiral armchair (left panel) 
and metallic zigzag (right panel) families. 
Similar fixed Rashba region lengths are considered, $L=4.92$ nm and $L=5.12$ nm 
for armchair and zigzag tubes, respectively.
}
\label{fig5}
\end{figure}

\subsection{Radius dependence}

As discussed above, an important parameter to be taken into account is the 
CNT radius. 
A detailed analysis of the dependence of the polarization on the tube radius 
is given in Fig. \ref{fig5} for the two families of achiral CNTs, 
$(n,n)$ armchair and $(n,0)$ metallic zigzag ($n=3q$) tubes. 
A color polarization map, $P_y (E,R)$, explicitly shows the spin polarization 
as a function of the energy and the tube radius. 
In the calculations we have considered similar Rashba region lengths to compare the 
spin polarization features in the two families. 
The figure shows how stronger polarization values are attained 
as the CNT radius increases, 
as well as how the maxima get closer to the Fermi energy. 
This is because for increasing radius, %when the radius increases, 
the second conduction band (valence band for negative energies) 
moves closer to the Fermi energy, providing the second channel needed 
for the obtention of a polarized spin current. 
Thus, for tubes with larger radii, spin polarization is more accessible. 
Note that the values of the spin-polarized current do not diminish in the 
plotted energy range, showing that the effect is robust with respect to 
increasing tube radius. This is at variance with the curvature-induced SOI, that diminishes 
for larger diameter nanotubes. Comparison to flat geometries, i.e., graphene nanoribbons \cite{Chico2015},
shows that flat and curved systems behave similarly under Rashba interaction for this particular 
projection direction, which maximizes the effect. Obvious differences arise due to the appearance of bands at different energies, 
but the maximum values of $P_y$ are of the same magnitude, and in both cases occur at the onset of new conductance channels. 

\begin{figure}[h]
\includegraphics[width=1.0\columnwidth]{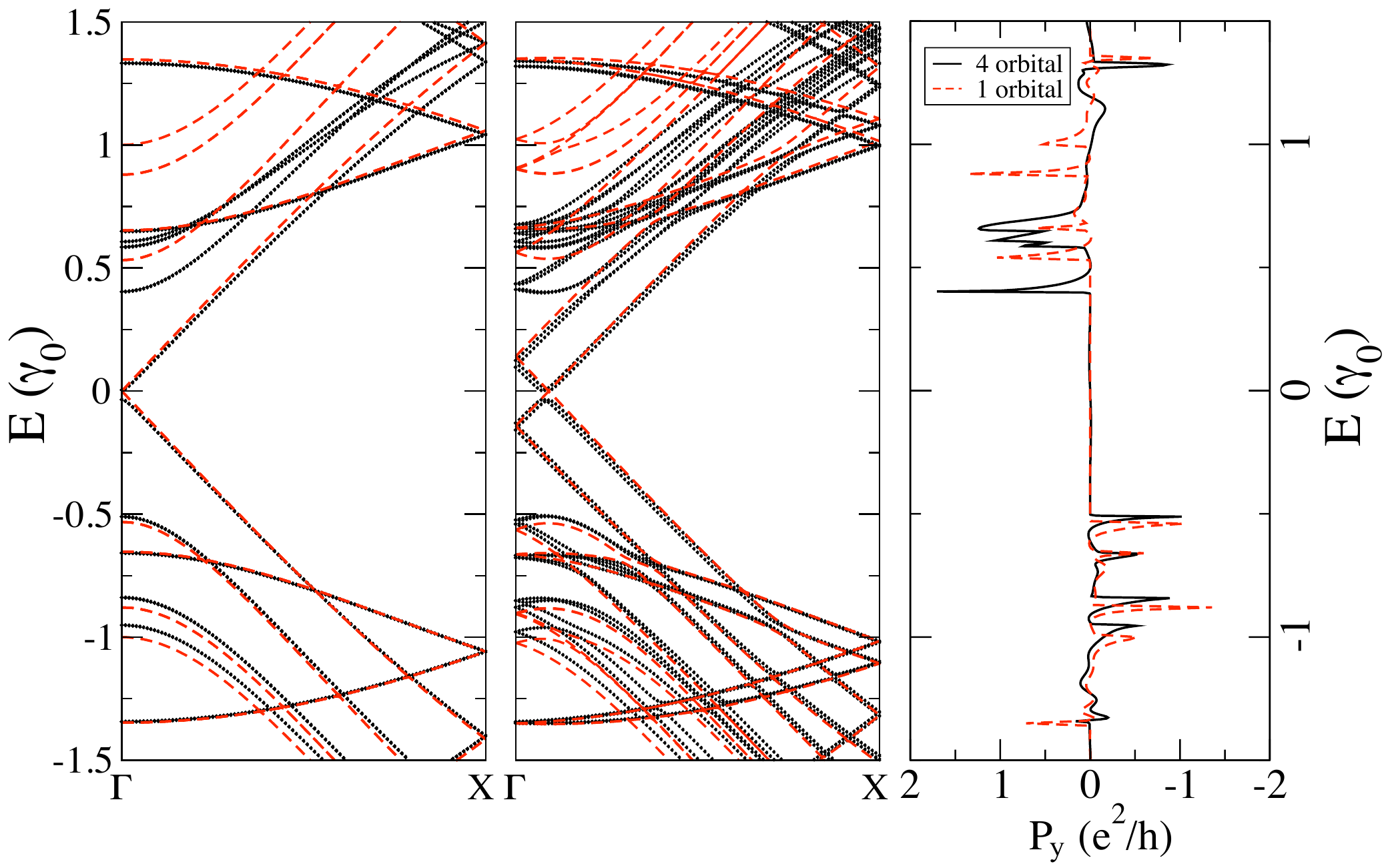}
\caption{(Color online) Electronic band structures for a (9,0) zigzag tube with Rashba region equal to 5 layers ($L=2,14$ nm), considering one (red dashed lines) and four orbitals (black curves) for $\lambda_R=0$ (left panel) and $\lambda =0.1 \gamma_0$ (middle panel). The right panel shows the spin-dependent current polarizations calculated with the two models. }
\label{fig6}
\end{figure}

\subsection{Multi-orbital model} 

In order to assess the previous results, we have compared to a more realistic 
tight-binding description.
We have calculated the band structure and the spin-polarized current of chiral 
CNTs described by a four-orbital model employing the Tom\'anek-Louie 
parameterization \cite{Tomanek88}. 
As in this model the tight-binding energy hoppings between two adjacent 
carbon atoms depend on their positions, the effect of the tube curvature is 
included.

The results for a (9,0) armchair CNT are shown in Fig. \ref{fig6}. 
Close to Fermi energy, leaving aside the gap opening, the electronic structure 
with a four-orbital model has no important changes with respect to the one-orbital 
model. 
However, some other differences appear, namely: 
(i) the electron-hole symmetry is broken, as expected; 
(ii) when compared with the one-orbital results, some bands move 
with respect to the Fermi level.

With respect to the spin polarization, the most remarkable difference 
in the energy interval studied here is the shift in energy and the height 
of the peaks or bumps. 
In any case, the spin polarization gap is smaller within the four-orbital 
model. 
This effect has been found for all chiralities, and 
not only for those presented in Fig. \ref{fig6}.

As we have corroborated the non-zero polarization results in the energy 
range of interest using the multi-orbital model with only some small 
quantitative differences, 
we conclude that the more realistic model validates the 
main features of the single-orbital approximation, 
that will be used in the rest of this work. 

\section{Symmetry considerations}
\label{sec:sym}
\subsection{Symmetry analysis of the spin-resolved conductances}

Up to this point, we have focused on the configurations that maximize 
the spin-polarized current, without discussing the symmetry considerations 
which could allow for the absence or existence of spin-polarized 
currents for all spin projections. 
In principle, one could expect that the same reasoning employed for 
planar systems \cite{Chico2015} should apply in this situation,  namely, 
to put the focus on symmetry operations 
{\it acting simultaneously on spatial and spin variables} 
that leave the system (the nanotube plus the direction of the electric field) 
invariant. 
This would be equivalent to the planar geometry studied in previous works \cite{Zhai2005,Chico2015}. 

In particular, if the nanotube unit cell is misaligned with respect to the 
electric field, so it does not coincide with a symmetry axis nor is included in any mirror plane, 
% that no axis of symmetry coincides with it
%nor mirror plane of the nanotube includes it, 
we expected no symmetry relations for the 
spin-resolved conductances. 
This was the case of the calculation presented in Fig. \ref{fig2} (a)
for an armchair (9,9) nanotube, 
because the applied electric field was not contained in any 
of the nanotube symmetry planes, the mirror plane $M_y$.
However, the results for the conductances found the difference 
$G_{\uparrow \downarrow}^{RL} - G_{\downarrow \uparrow}^{RL}$ 
to be exactly zero (within our numerical accuracy)
for all rotation angles $\phi$ around the nanotube axis, i.e.,  
no matter if the mirror symmetry $M_y$ is present or not.

\begin{figure}[t]
\includegraphics[width=0.95\columnwidth]{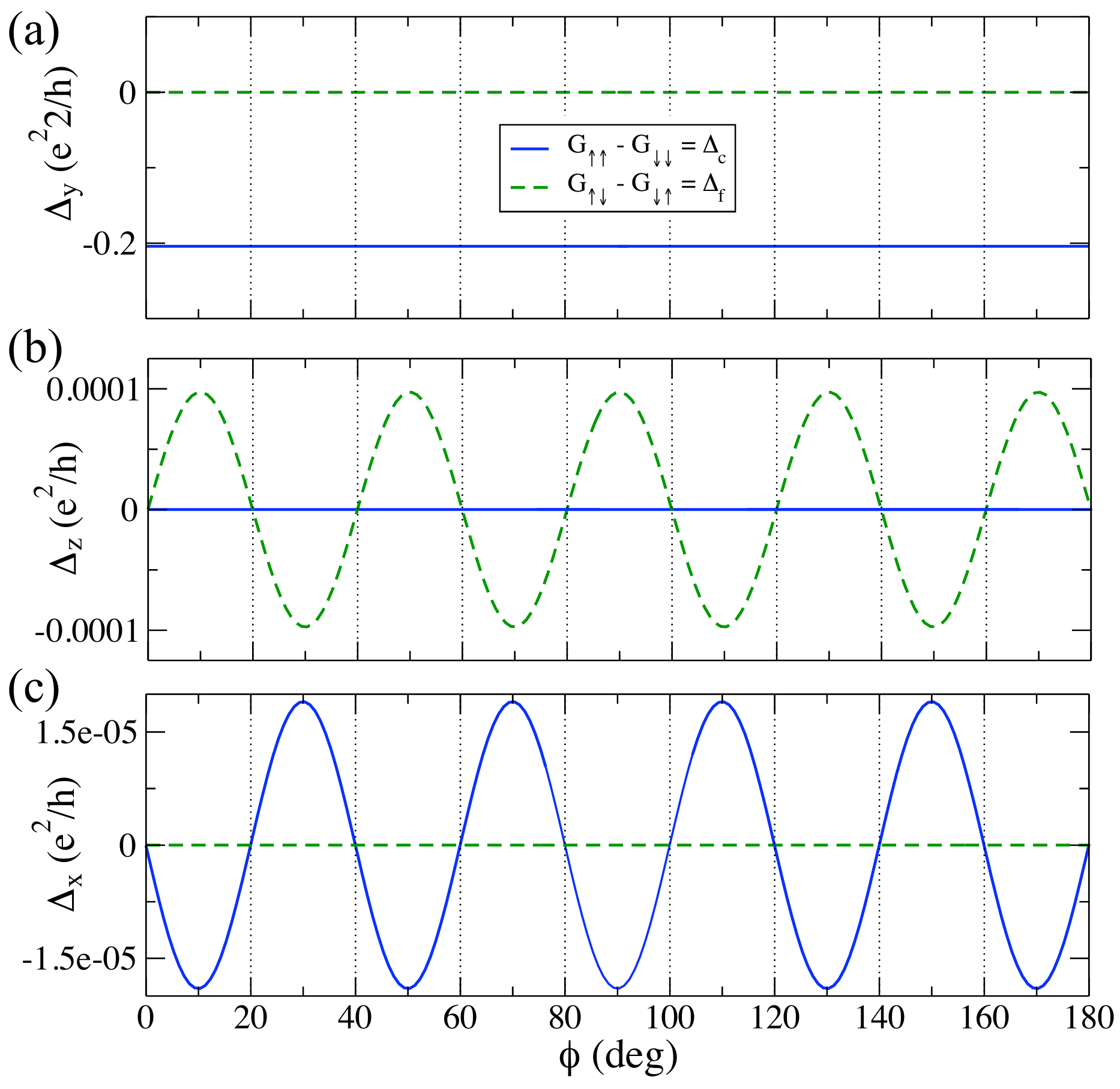}
\caption{(Color online) 
Difference between projected spin-conserved and spin-flip conductances, 
$\Delta_c$ and $\Delta_f$, as functions of the rotation angle for a 
armchair (9,9) CNT with Rashba region equal to 16, for a fixed energy $E=-0.983 \gamma_0$. The differences for the three spin projection directions $y, z$ and  $x$ are labeled $\Delta_y$, $\Delta_z$, and $\Delta_x$, respectively.}
\label{cnt99rot}
\end{figure}

This assertion is illustrated in Fig. \ref{cnt99rot} (a), in which we show the difference 
between the spin-conserved 
$\Delta_c=G_{\uparrow\uparrow}^{LR} - G_{\downarrow\downarrow}^{LR}$ 
(blue curve) 
and spin-flip 
$\Delta_f=G_{\uparrow\downarrow}^{LR} - G_{\downarrow\uparrow}^{LR}$ 
(green dashed curve) 
conductances in the $y$ direction for a (9,9) CNT with
a Rashba region of 16 unit cells as a function of  $\phi$, at a fixed energy 
$E=-0.983\gamma_0$. 
The spin-flip conductances are equal and independent of the rotation angle 
for this spin projection direction. 
This equality points towards the existence of at least one symmetry that we have 
not considered yet. 
We should therefore look for more general symmetries, such as those operating independently in spatial and spin spaces \cite{Bhatta2013,Bhatta2014}. On the other side, the values of $\Delta_c$ or $\Delta_f$  
for other spin-projection directions oscillate, 
as presented in panels (b) and (c) of Fig. \ref{cnt99rot}. 
We must remark that the amplitude of these oscillations are much smaller than 
the difference between $\Delta_c$ and $\Delta_f$ in the $y$ direction.

%\begin{center}
%\end{center}

\begin{figure}[t]
\includegraphics[width=1.\columnwidth]{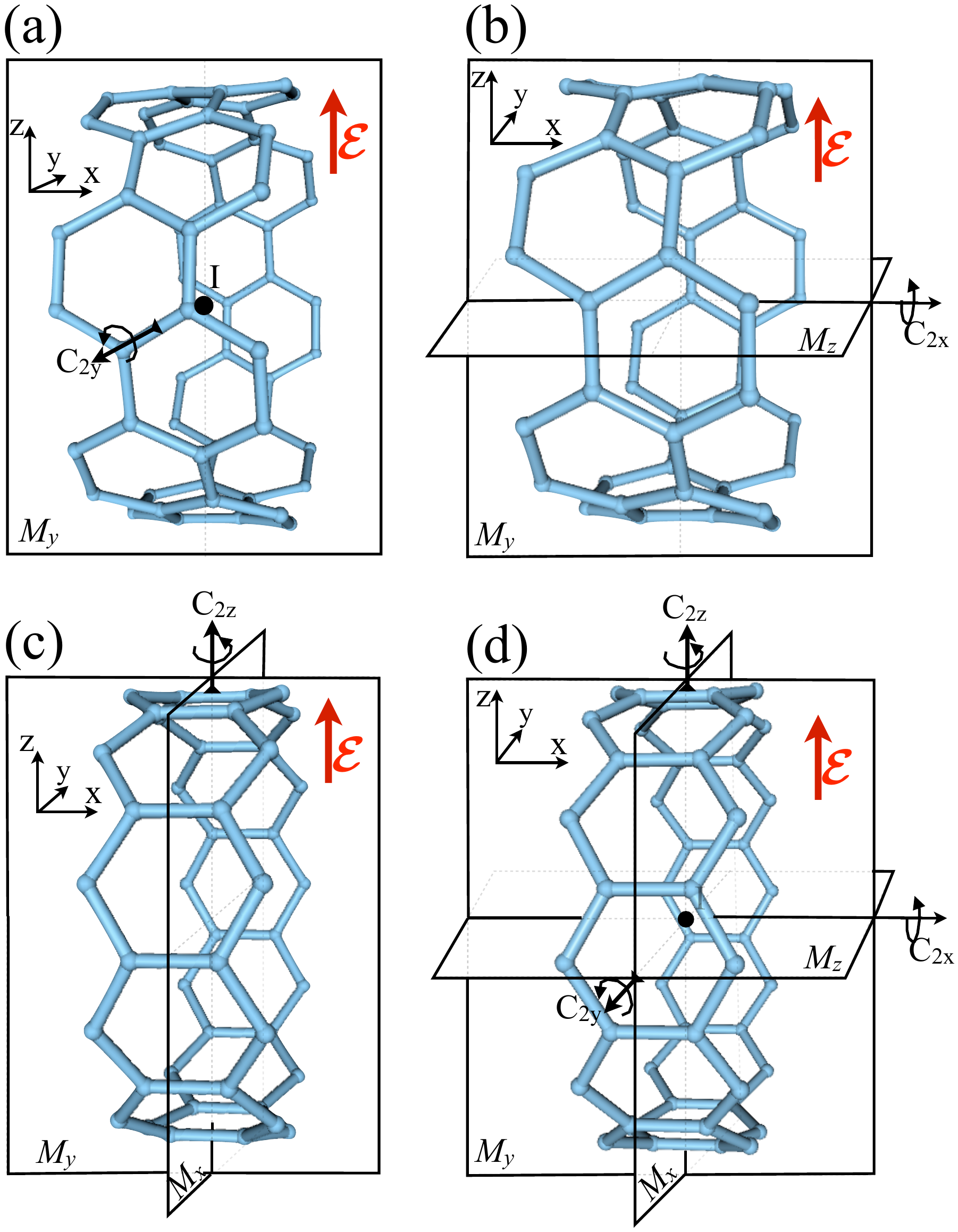}
\caption{(Color online)  
Schematic plot of CNT unit cells with the symmetry mirror planes, rotation axes  and inversion center $I$---when applicable---for different achiral tubes: (a)  (7,7) and (b)  (6,6) armchair CNTs, and (c) (11,0) and (d) (12,0) armchair tubes. The mirror planes with normal vectors in $x$, $y$, and $z$ directions are labeled $M_{x}$, $M_{y}$, and $M_{z}$, respectively. The rotation symmetry operations, named as $C_{2\,x,y,z}$, are also indicated.}
\label{symops}
\end{figure}

Fig. \ref{symops} shows the unit cells of the four instances of achiral carbon 
nanotube unit cells with different symmetries, $(n,n)$ and $(n,0)$, 
with $n$ odd and even. 
The unit cells are oriented in  a symmetrical fashion with respect to the electric field 
$E_z$.
We first discuss  the armchair (7,7) CNT, which has 
the same symmetry operations as the previously discussed (9,9) case. 
We detail the three main symmetries of this nanotube, of interest for the Rashba Hamiltonian 
(from now on we use $(r)$ and $(s)$ superscripts to label a symmetry operation 
performed in real and spin spaces, respectively).
Firstly, we notice in Fig. \ref{symops} (a) a mirror plane, $xz$; the corresponding mirror reflection operation in spatial coordinates is labeled $M_y^{(r)}$. 
Associated to this $M_y^{(r)}$ symmetry there is a perpendicular $C_2$ axis in the $y$ direction. This operation is denoted $C_{2y}^{(r)}$; it changes the sign of the spatial part of the Rashba term. 
In general, an $(n,n)$ CNT in an electric field along $z$, with $n$ odd, has $n$ mirror planes that coincide with the electric field when 
rotating the tube around its axis; 
these symmetry planes would yield an oscillating pattern in the conductance 
differences as a function of the rotation angle $\phi$, as seen in 
Fig. \ref{cnt99rot} (b) and (c).
It should also be noted that in the absence of an electric field, the spatial inversion $I^{(r)}$ leaves the nanotube invariant 
for any rotation angle. 

We can combine spatial and spin symmetry operations to keep invariant the Rashba Hamiltonian.
With respect to the inversion operation, here is a change of sign of $H_R$ under spatial 
inversion $I^{(r)}$. 
This is better visualized by resorting to the equivalent continuum form, 
$H_R \propto (\sigma_x k_y -\sigma_y k_x)$.
By performing a spatial inversion $I^{(r)}$ and a rotation of 180$^{\rm o}$ in spin space 
around the $z$ axis, $C_{2z}^{(s)}$, the Rashba Hamiltonian remains invariant. 
When the $M_y^{(r)}$ symmetry is present, 
we need a $C_{2y}^{(s)}$ rotation in order to leave the Hamiltonian 
invariant; and 
finally, the $C_{2y}^{(r)}$ axis changes the sign of the Rashba term, so 
a related $C_{2x}^{(s)}$ rotation in spin space is needed.

Let us now analyze the relations that can be inferred from the symmetry $I^{(r)} \otimes C_{2z}^{(s)}$ upon the spin-resolved conductances: 
 spatial inversion $I^{(r)}$ yields ($x,y,z$)$ \ \rightarrow \ $($-x,-y,-z$),
%  ($p_{x}$,$p_{y}$,$p_{z}$)$ \ \rightarrow \ $($-p_{x}$,$-p_{y}$,$-p_{z}$), 
  whereas $C_{2z}^{(s)}$ implies the spin transformation ($\sigma_{x}$,$\sigma_{y}$,$\sigma_{z}$)$ \ \rightarrow \ $($-\sigma_{x}$,$-\sigma_{y}$,$\sigma_{z}$). So if the spin projection direction is along $x$ or $y$, 
we have $G^{LR}_{\sigma \sigma '} = G^{RL}_{\bar {\sigma} \bar{\sigma} '}$. 
Then, considering time-reversal symmetry $\Theta$, 
$G^{LR}_{\sigma \sigma '} = G^{RL}_{\bar { \sigma}' \bar{\sigma}}$, 
we have $G^{LR}_{\sigma \sigma '} = G^{LR}_{\sigma' \sigma } $, i.e., $G_{\uparrow\downarrow}^{LR} =  G_{\downarrow\uparrow}^{LR}$, as we obtained numerically 
(remember Figs. \ref{cnt99rot} (a) and (c), which have $\Delta_f=0$ for all $\phi$). 
For the $z$ spin direction, we have 
 $G^{LR}_{\sigma \sigma '} = G^{RL}_{\sigma \sigma '}$, that leads to  $G^{LR}_{\uparrow\uparrow} =  G^{LR}_{\downarrow\downarrow}$ for all tube orientations $\phi$. 
 This is also clearly seen in Fig. \ref{cnt99rot} (b), which shows $\Delta_c=0$ for all $\phi$. 

%\begin{center}
%\end{center} 

As we already mentioned, besides the symmetry $I^{(r)} \otimes C_{2z}^{(s)}$, 
 always present in $(n,n)$ tubes with odd $n$, irrespectively of the CNT orientation, 
there are other symmetries which appear for certain rotation angles, that is, 
every $360^{\rm o}/n$. 
These are $M_y$ acting in real and spin spaces, 
that can be also written as $M_y^{(r)} \otimes C_{2y}^{(s)}$, and the spatial rotation $C_{2y}^{(r)}$ complemented with the rotation in spin space  $C_{2x}^{(s)}$, i.e., 
$C_{2y}^{(r)} \otimes C_{2x}^{(s)}$. 
These two symmetries occur simultaneously. 
It can be verified that they yield a zero spin polarization when the electric field 
coincides with the mirror plane, but not so for intermediate angles, giving rise to 
the oscillating behavior of $\Delta_f$ for spin projection in the $z$ direction and 
that of $\Delta_c$ for spin in the $x$ direction, as it can be seen in 
Figs.  \ref{cnt99rot} (b) and (c), respectively. 

We can apply this line of reasoning to all the possible symmetries present in 
finite-size carbon nanotubes and spin projection directions.
Notice that not all the symmetry operations of the finite-size unit cell in the absence 
of an electric field are valid; only those that either leave the electric field term invariant 
or with a minus sign change are acceptable. 
This rules out improper rotations $S_{2n}$ in achiral nanotubes. 
In the Appendix we give a list of the symmetries of finite-sized CNTs. 
 
In Fig. \ref{symops} we present all the symmetry operations in real space for 
achiral CNTs for all even and odd indices, 
with a Rashba field that either leave the Rashba term invariant or produce a sign 
change in it (we have shown how to compensate with a rotation in spin space).
For example, for a zigzag $(n,0)$ CNT with $n$ even (Fig.  \ref{symops} (c)), we can find a mirror $M_x$ operation 
acting in both spatial and spin spaces, i.e., $M_x^{(r)} \otimes C_{2x}^{(s)}$, 
but also a $C_{2y}^{(r)}$ sign-changing rotation appearing  every $360^{\rm o}/n$, 
which leads to the symmetry $C_{2y}^{(r)} \otimes C_{2x}^{(s)}$. 

All these symmetries have been included in Table \ref{tablesym}, 
where we show the final outcome for the conductance relations and 
the values of spin polarizations, invoking time-reversal symmetry when needed. 
This is necessary when the spatial symmetry interchanges the role of the left and 
right electrodes.

 \begin{table}[t]
\renewcommand{\arraystretch}{2}
\setlength{\tabcolsep}{2pt} 
\caption{Symmetry, conductance and polarization relations for nanotube systems with Rashba spin-orbit interaction.}
\begin{center}
\begin{tabular}{|c|c|c|c|}
      \hline
   {Symmetries}   &  {Conductance} & \multicolumn{2}{c|}{Spin polarization} \\  \cline{3-4}
%      ($r \otimes s$)& $(x,y,z)$  & $G_{\uparrow\uparrow}-G_{\downarrow\downarrow}$  & %$G_{\uparrow\downarrow}-G_{\downarrow\uparrow}$ \\
 ($r \otimes s$)& $(x,y,z)$  & $G_{\uparrow\uparrow}-G_{\downarrow\downarrow}$  & $G_{\uparrow\downarrow}-G_{\downarrow\uparrow}$ \\
 %$\Delta_c$  & $\Delta_f$ \\
      \hline
%       & \multirow{2}{*} {$(x,z)$  $G^{LR}_{\sigma\sigma'} = G^{LR}_{\bar\sigma\bar\sigma'}$} & \multirow{2}{*}{0} & \multirow{2}{*}{0} \\ 
%%        {$C^{(r)}_{2x} \otimes C^{(s)}_{2y}$} & {}Ê& {}  &  {} \\ \cline{2-4}
%  {$C^{(r)}_{2x} \otimes C^{(s)}_{2y}$} & {}& {}  &  {} \\ \cline{2-4}
%  $M^{(r)}_{y} \otimes C^{(s)}_{2y}$   & \multirow{2}{*}{ $(y)$  $G^{LR}_{\sigma\sigma'} = G^{LR}_{\sigma\sigma'}$} & \multirow{2}{*}{$\neq0$} &\multirow{2}{*}{$\neq0$} \\
%    &   &    &     \\  
%   \hline
       \hline
        {$C^{(r)}_{2x} \otimes C^{(s)}_{2y}$} & $(x,z)$  $G^{LR}_{\sigma\sigma'} = G^{LR}_{\bar\sigma\bar\sigma'}$ & $0$ & $0$ \\ \cline{2-4}
    $M^{(r)}_{y} \otimes C^{(s)}_{2y}$   &  $(y)$  $G^{LR}_{\sigma\sigma'} = G^{LR}_{\sigma\sigma'}$ & $\neq0$ & $\neq0$ \\  
     \hline
%        \multicolumn{1}{|c|}{\multirow{2}{*}{$C^{(r)}_{2y} \otimes C^{(s)}_{2x}$ and $C^{(r)}_{2z} \otimes C^{(s)}_{2z}$} } & $(y,z)$ $G^{LR}_{\sigma\sigma'} = G^{LR}_{\sigma'\sigma}$ & $\neq0$ & $0$ \\ \cline{2-4}
%       &  $(x)$ $G^{LR}_{\sigma\sigma'} = G^{LR}_{\bar\sigma'\bar\sigma}$ & $0$ & $\neq 0$  \\ 
%     \hline
      {$I^{(r)} \otimes C^{(s)}_{2z}$}   & $(x,y)$ $G^{LR}_{\sigma\sigma'} = G^{LR}_{\sigma'\sigma}$ & $\neq 0$ & $0$ \\ \cline{2-4}
  {$C^{(r)}_{2z} \otimes C^{(s)}_{2z}$}     & $(z)$ $G^{LR}_{\sigma\sigma'} = G^{LR}_{\bar\sigma'\bar\sigma}$ &  $0$ & $\neq0$ \\ 
     \hline
%             \multicolumn{1}{|c|}{\multirow{2}{*}{$I^{(r)} \otimes C^{(s)}_{2z}$}} & $(x,y)$ $G^{LR}_{\sigma\sigma'} = G^{LR}_{\sigma'\sigma}$ & $\neq 0$ & $0$  \\ \cline{2-4}
%       & $(z)$ $G^{LR}_{\sigma\sigma'} = G^{LR}_{\bar\sigma'\bar\sigma}$ & $0$ & $\neq0$   \\ 
%            \hline
 $C^{(r)}_{2y} \otimes C^{(s)}_{2x}$ & $(y,z)$ $G^{LR}_{\sigma\sigma'} = G^{LR}_{\sigma'\sigma}$ & $\neq 0$ & $0$ \\ \cline{2-4}
  $M^{(r)}_{x} \otimes C^{(s)}_{2x}$       &  $(x)$ $G^{LR}_{\sigma\sigma'} = G^{LR}_{\bar\sigma'\bar\sigma}$ & $0$ & $\neq0$  \\   
      \hline
%       \multicolumn{1}{|c|}{\multirow{2}{*}{$M^{(r)}_{y} \otimes C^{(s)}_{2y}$}} & $(x,z)$ $G^{LR}_{\sigma\sigma'} = G^{LR}_{\bar\sigma\bar\sigma'}$ & $0$ & $0$  \\ \cline{2-4}
%       & $(y)$ $G^{LR}_{\sigma\sigma'} = G^{LR}_{\sigma\sigma'}$ & $\neq0$ & $\neq0$ \\   
%\hline
\end{tabular}
\end{center}
\label{tablesym}
\end{table}
 
Notice that in Table \ref{tablesym} we have grouped the symmetries which give 
rise to the same conductance relations; they happen to appear in pairs. 
For each case, the first row corresponds to a non-trivial symmetry, in the sense 
that different operations are performed in spatial and spin variables. 
These are $C_{2x}^{(r)} \otimes C_{2y}^{(s)}$, $I^{(r)} \otimes C_{2z}^{(s)}$, 
and $C_{2y}^{(r)} \otimes C_{2x}^{(s)}$, which are necessary to explain the 
CNT results. 
The second row corresponds to symmetries that were already discussed in 
the planar case: for example, $M_x^{(r)} \otimes C_{2x}^{(s)}$ is just the mirror 
reflection $M_x$ considered both in real and spin spaces, and a rotation has 
trivially the same effect in spin and spatial variables. 
Planar systems such as graphene ribbons may also have some of these 
non-trivial symmetries. 
We have checked that they are consistent with our previous results, yielding 
the same conductance relations already reported in 
Ref. \onlinecite{Chico2015}.

\begin{figure}[t]
\includegraphics[width=0.95\columnwidth]{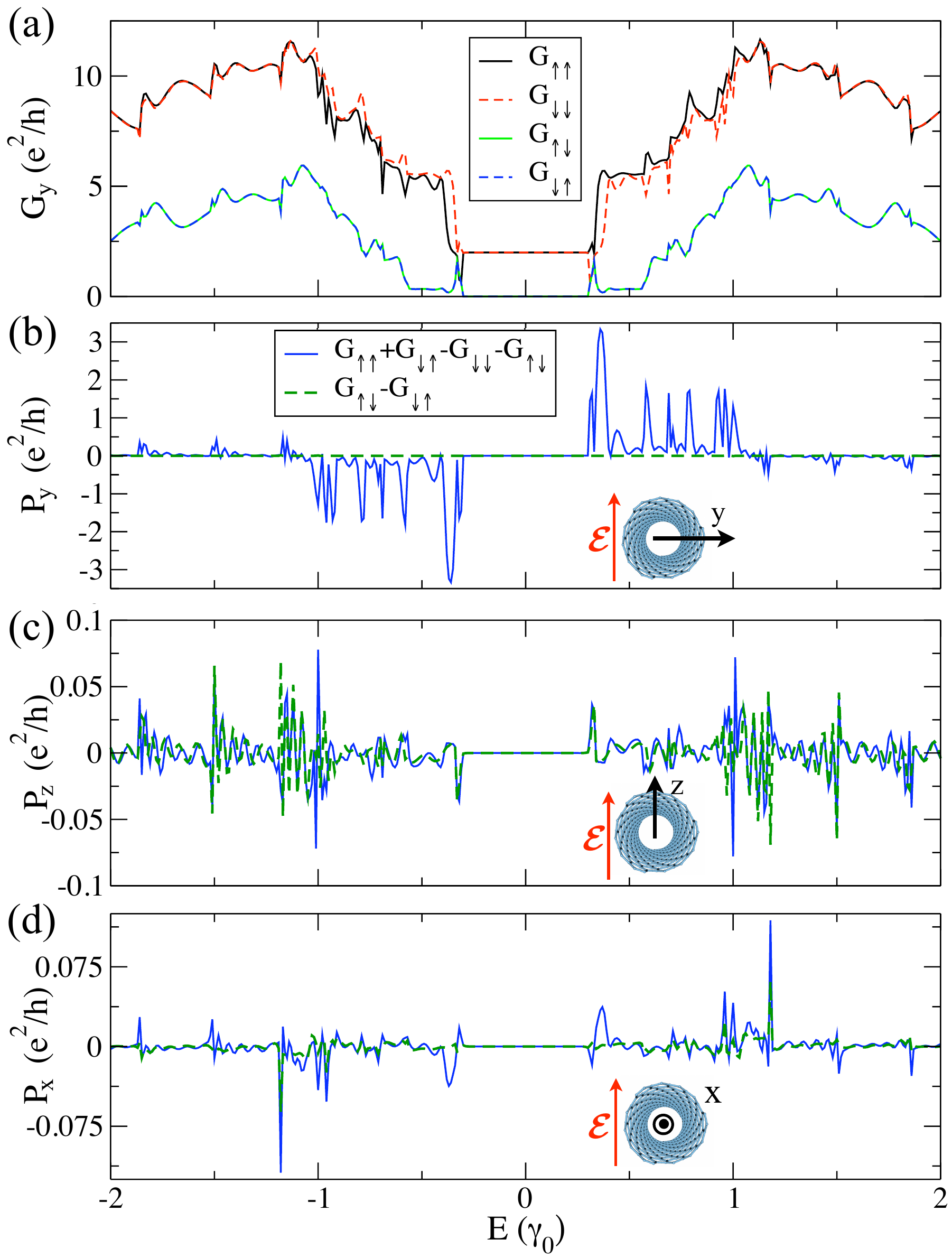}
\caption{(Color online) (a) Spin-dependent conductances for a (15,3) CNT with Rashba region equal to $6$  and considering the spin on the ${y}$ direction. (b), (c), and (d) correspond to spin polarization of the conductances, with spin projection direction in the ${y}$, $z$, and ${x}$ directions, respectively, shown in the insets. Note the different polarization scales used in each panel.}
\label{fig153}
\end{figure}

%\begin{center}
\begin{figure}[t]
\includegraphics[width=.95\columnwidth]{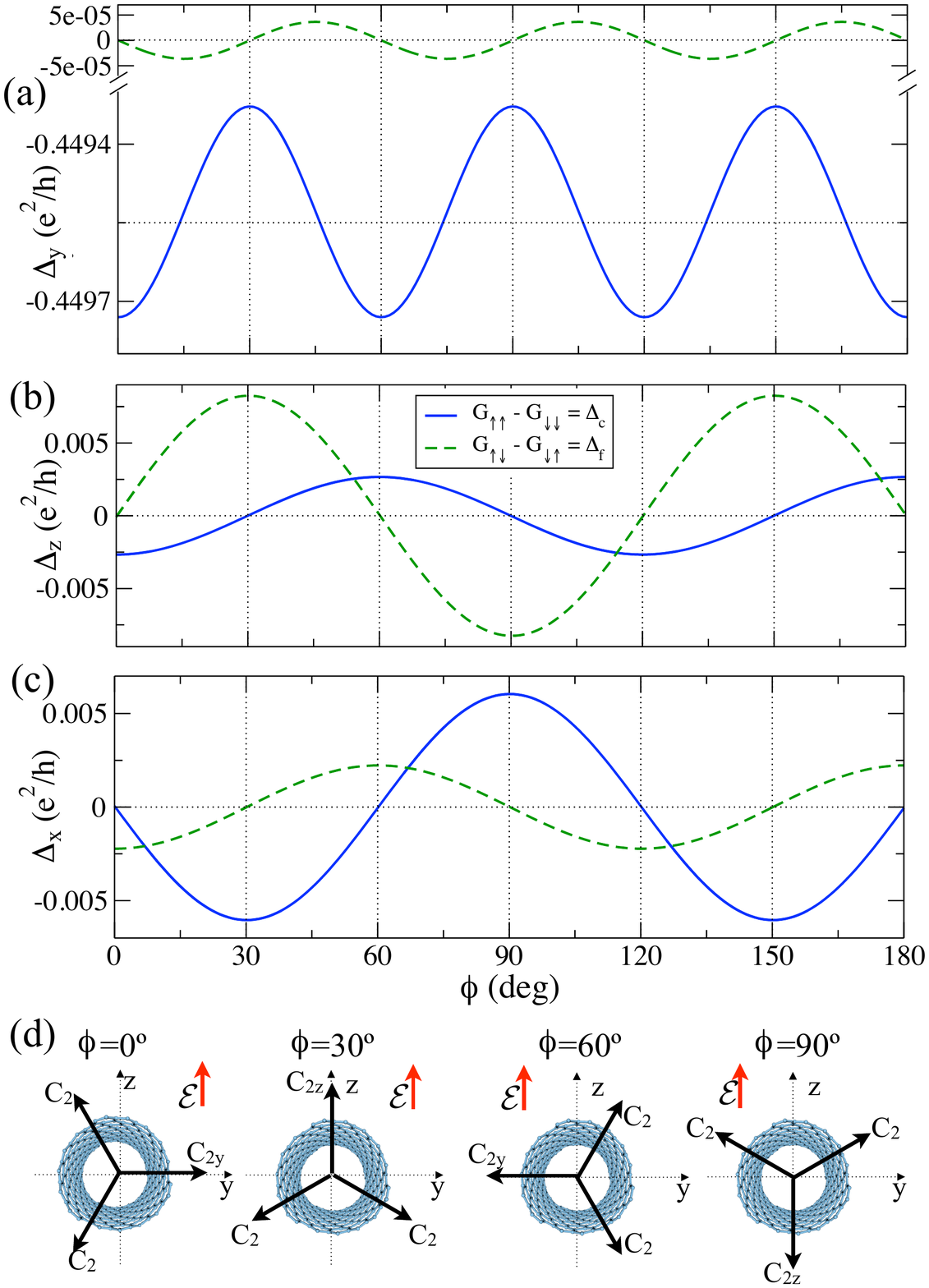}
\caption{(Color online)  Difference between spin-conserved and spin-flip conductances as functions of the rotation angle for a (15,3) nanotube with Rashba region equal to $6$, for a fixed energy $E= -0.425 \gamma_0$.  The $y, z$, and $x$ spin projection directions are considered in panels (a), (b), and (c), respectively. (d) Schematic drawings of the  three $C_2^{(r)}$ axis for different rotation angles.}
\label{rot153}
\end{figure}
%\end{center}

\subsection{Chiral carbon nanotubes}

Chiral nanotubes have fewer symmetries than the achiral ones, 
especially for the finite-size regions considered in this work. 
If the nanotube indices $(n,m)$ have a common divisor $p$, then
the  CNT is invariant under $360^{\rm o}/p$ rotations around the tube 
axis, $C_{px}^{(r)}$, leaving the Rashba term invariant or changed by a 
minus sign depending both on the value of $p$ and the CNT orientation 
with respect to the electric field. 
It has also $p$ $C_{2}^{(r)}$ rotation axes perpendicular to the CNT axis $x$.

In order to illustrate these symmetries with an example, we choose the (15,3) CNT. 
This nanotube has three $C_2^{(r)}$ axes perpendicular to the tube at 120$^{\rm o}$. 
Fig. \ref{fig153} shows in panel (a)  the spin-dependent conductances in the $y$ 
direction,  and the current polarization of the (15,3) tube for the three spin 
projection directions in panels (b-d), considering the same Rashba region with $N=6$.  
Differently from the results found for the achiral tubes, net polarized currents are now possible for the three spin directions, $x$, $y$, and $z$. 
But as already discussed in the previous Section, the greatest spin-polarized 
current is obtained in the $y$ direction, 
perpendicular to the applied field and to the tube axis. 
The values are two orders of magnitude larger than in the $x$ and $z$-direction, 
and of the same order of magnitude as those observed for achiral tubes;  
see Fig. \ref{fig2}.

We see in panels (b) and (c) of Fig.  \ref{fig153} that in principle no relation 
between the spin-flip and spin-conserved conductances hold; however, 
the spin-flip term seems to be zero for the $y$ spin projection direction 
shown in panel (a). 
 In order to clarify this, we proceed as in the armchair case. We pick a fixed 
 energy and study the dependence of the conductances on the rotation angle 
 $\phi$ around the CNT axis. 
 The differences in the spin-flip and spin-conserved conductances as functions 
 of the rotation angle for the three spin projection directions are shown in 
 Fig. \ref{rot153}. 
When compared with the results presented in Fig. \ref{cnt99rot}, 
we see that all $\Delta_c$ and $\Delta_f$ along the $x$, $y$ and $z$ directions 
oscillate, and for the $x$ and $z$ directions the values are about two orders of 
magnitude greater. 
 
The most conspicuous feature of the conductance differences $\Delta_c$ and 
$ \Delta_f$ is that the oscillation period is different for the $y$ spin projection 
direction, being twice the period of $x$ or $z$ directions. 
From the schematic drawings at the bottom of Fig. \ref{rot153}, we see that 
at every 30$^{\rm o}$ there is one $C_2^{(r)}$ axis either aligned with the $z$ 
or with the $y$ direction, so that the symmetry of the nanotube oscillates 
between 
$C_{2y}^{(r)} \otimes C_{2x}^{(s)}$ and $C_{2z}^{(r)} \otimes C_{2z}^{(s)}$. 
This oscillation implies that $\Delta_c$, $\Delta_f$ are zero and non-zero 
alternatively every 30$^{\rm o}$ with varying $\phi$ for the $x$ and $z$ 
spin projection directions, but it gives $\Delta_f=0$ for $\phi=0^{\rm o}$, 
30$^{\rm o}$, 60$^{\rm o}$... 
for spin projection direction $y$, giving twice the period. 

We should emphasize that the symmetry analysis gathered in Table \ref{tablesym} 
can be applied to any CNT with a finite-size RSO interaction in order to 
predict the occurrence of spin-polarized currents in different directions, 
and also to explain all the conductance 
conservation relations observed in our calculations. 
We also expect that the analysis of spatial and spin symmetries will be of 
interest for the study of spintronic devices, and additionally in other problems 
where the interplay of spin and spatial variables is relevant.

%\begin{table}[ht]
%\renewcommand{\arraystretch}{2}
%\setlength{\tabcolsep}{2pt}
%\caption{Summary of the Symmetries in the tubes}
%\begin{center}
%\begin{tabular}{|c|c|c|c|}
%%\begin{tabular}{cccc}
%%\begin{tabular}{ccc}
%%   \hline
%     \hline
%   \multirow{2}{*}{CNT}   &  \multirow{2}{*}{Point group} & \multicolumn{2}{c|}{Symmetries} \\  \cline{3-4}
%     &   & $n$ even  & $n$ odd \\
%%   \hline
%      \hline
%       \multicolumn{1}{|c|}{\multirow{2}{*}{($n,n$)}} & \multicolumn{1}{c|}{\multirow{2}{*}{$D_{nd}$}} & $C^n_{x}$, $nC^2_{\perp x}$, $nM_{\perp x}$,  & $C^n_{x}$, $nC^2_{\perp x}$, $nM_{\perp x}$,  \\
%      &   &  $nS^{2n}$ & $I$, $nS^{2n}$ \\
%    \hline
%       \multicolumn{1}{|c|}{\multirow{2}{*}{($n,0$)}} & \multicolumn{1}{c|}{\multirow{2}{*}{$D_{nh}$}} & $C^n_{x}$, $nC^2_{\perp x}$, $nM_{\perp x}$,  & $C^n_{x}$, $nC^2_{\perp x}$, $nM_{\perp x}$, \\
%      &   & $I$, $M_x$, $nS^{2n}$ & $M_x$, $nS^{2n}$  \\
%    \hline
%       \multicolumn{1}{|c|}{\multirow{2}{*}{($n,m$)}} & \multicolumn{1}{c|}{\multirow{2}{*}{$D_N$}} & \multicolumn{2}{c|}{\multirow{2}{*}{$C^N_{x}$, $NC^2_{\perp x}$}}  \\
%       &  & \multicolumn{2}{l|}{{\tiny \it{N:mcd(n,m)}}}    \\
%    \hline
%\end{tabular}
%\end{center}
%\label{table}
%\end{table}
%

\section{Summary and Conclusions}
\label{sec:sum}

We have presented a detailed study of spin-resolved conductances and spin 
polarization currents in carbon nanotube systems of different chiralities and 
under the effect of Rashba SOI in a finite part of the tube.  
Similarly to graphene nanoribbons, we have shown that the best geometry 
to achieve a spin-polarized current in CNTs is when the electric field, the spin 
polarization direction and the tube axis are all perpendicular to each other.

However, there are important differences with respect to planar systems.  
In the latter, such as graphene ribbons, it is sufficient to consider symmetry 
operations acting simultaneously on spatial and spin variables in order to 
predict the existence of spin-polarized currents in different spin directions. 
But, as shown in this paper, more general symmetries have to be 
invoked to provide a correct description of carbon nanotubes. 
%correctly describe each one of the chiral and achiral tubes. 

We have demonstrated that different symmetries %operating independently
acting independently in spatial and spin 
%spaces 
variables of the Rashba CNT Hamiltonian are needed 
%used
to explain and predict the spin transport behavior of these systems. 
%guarantee the invariance of the Hamiltonian of the tube with the electric field.
A full understanding of the relevant symmetries allows us
to elucidate the characteristics of the spin-dependent conductance 
and polarization of CNTs.
Furthermore, we expect this symmetry analysis to be valuable for the 
understanding of other systems for which spin and spatial variables are 
related in a non-trivial way.

Our results open the possibility for the design of 
an all-electrical spin valve based in CNTs driven by Rashba coupling, provided that its value is enhanced by proximity or hybridization effects.  %in different directions. 
All these findings may be applied to other materials with Rashba spin-orbit interaction, serving a guide to maximize the spintronic response of the devices.

%As an example, we compute the spin-dependent transport of graphene nanoribbons with an applied electric field in finite region. We have shown that spin-polarized currents can be achieved if the spin polarization is measured in the transversal direction of the ribbon for all the ribbon geometries.  Furthermore, we have analyzed 
%all the basic symmetries and spin directions, elucidating which configurations can yield a spin-polarized current on the basis of symmetry.  
%The intensity and sign of the Rashba spin-orbit coupling may be modified by external electric field, opening the possibility of building an all-electrical spin valve.
%Our findings can be useful for a smart design of spintronic graphene devices, being of general application to other materials with Rashba SOI.  
%in other to elucidate the geometries of interest for the obtention of spin-polarized currents.  

\section*{acknowledgments}
We  thank L. Brey for interesting discussions. This work has been financially supported by  FAPERJ under grant E-26/102.272/2013 and by Spanish MINECO grants No. FIS2012-33521 and FIS2015-64654-P. We acknowledge the financial support of the CNPq/CSIC project 2011BR0087 and of the INCT de Nanomateriais de carbono.

\appendix*

\section{}

In this Appendix we give a summary of the spatial symmetries of 
finite-sized CNTs. 
For simplicity, we assume the unit cell to be of the most symmetrical form,
so that the symmetries are in fact those of the point group of the 
infinite CNT \cite{Damnjanovic1999,Alon2000}. 
For achiral tubes, the chosen unit cells are those shown in Fig. \ref{symops}. 
For chiral CNTs, we pick a maximally 
symmetric unit cell, with the symmetries of the infinite CNT point group. 
Note that any differences in the results of the conductances 
due to the choice of unit cells will clearly diminish for 
large $N$, i.e., large Rashba regions. 

\begin{table}[!h]
\renewcommand{\arraystretch}{2}
\setlength{\tabcolsep}{2pt}
\caption{Spatial symmetries of a finite-size CNTs composed of an integer number $N$ unit cells. 
%For achiral tubes we choose the unit cells as depicted in Fig. \ref{symops}; chiral tubes are assumed to have a maximally symmetric 
%unit cell. 
Improper rotations, that do not play any role 
for the Rashba Hamiltonian considered in this work, are marked in red. }

\begin{center}
\begin{tabular}{|c|c|c|c|}
%\begin{tabular}{cccc}
%\begin{tabular}{ccc}
%   \hline
     \hline
   \multirow{2}{*}{Finite}   &  \multirow{2}{*}{Point} & \multicolumn{2}{c|}{Symmetry operations} \\  \cline{3-4}
    CNT & group  & $n$ even  & $n$ odd \\
%   \hline
      \hline
       \multicolumn{1}{|c|}{\multirow{2}{*}{$N(n,n$)}} & \multicolumn{1}{c|}{\multirow{2}{*}{$D_{nd}$}} & $C_{nx}$, $nC_{2\perp x}$, $nM_{\alpha \perp x}$,  & $C_{nx}$, $nC_{n\perp x}$, $nM_{\alpha \perp x}$,  \\
      &   &  {\color{red}$nS_{2n}$} & $I$, {\color{red}$nS_{2n}$} \\
    \hline
       \multicolumn{1}{|c|}{\multirow{2}{*}{$N(n,0$)}} & \multicolumn{1}{c|}{\multirow{2}{*}{$D_{nh}$}} & $C_{nx}$, $nC_{2\perp x}$, $nM_{\alpha \perp x}$,  & $C_{nx}$, $nC_{2\perp x}$, $nM_{\alpha \perp x}$, \\
      &   & $I$, $M_x$, {\color{red}$nS_{2n}$} & $M_x$, {\color{red}$nS_{2n}$} \\
    \hline
       \multicolumn{1}{|c|}{\multirow{2}{*}{$N(n,m$)}} & \multicolumn{1}{c|}{\multirow{2}{*}{$D_p$}} & \multicolumn{2}{c|}{\multirow{2}{*}{$C_{px}$, $p \ C_{2\perp x}$ \quad 
       $p = \gcd(n,m)$}}  \\
       &  & \multicolumn{2}{l|}{{}}    \\
    \hline
\end{tabular}
\end{center}
\label{tabletub}
\end{table}

Table \ref{tabletub} present the symmetry operations for a finite-sized nanotube  
with tube axis aligned with the $x$ direction.
We use the following nomenclature: 
$C_{nx}$ is a $360^{\rm o}/n$ rotation around the $x$ axis; 
for even $n$ the tube always presents a $C_{2x}$ rotation.
$nC_{2\perp x}$ corresponds to $n \ C_2$ rotations perpendicular 
to the tube axis ($x$).
$nM_{\alpha \perp x}$ specifies $n$ mirror symmetry planes 
$M_\alpha$, where $\alpha$ is an axis perpendicular to $x$. 
If $\alpha$ points in the $y$ or $z$ directions a $M_y$ or $M_z$ symmetry 
operation results, respectively.
$I$ is the inversion operation and $S_{2n}$ corresponds to 
improper rotations.

%\bibliography{CNTrso}

\begin{thebibliography}{57}
\bibitem [{Lch()}]{Lchiral}%
  \BibitemOpen
  \href@noop {} {}\bibinfo {note} {For chiral tubes $(n,m)$, the length of the
  unit cell is given by $T=\frac{3 a_c \sqrt{n^2 + m^2 + nm }}{d_R}$, where
  $d_R$ is the greatest common divisor of $2n+m$ and $2m+n$,
  $\gcd(2n+m,2m+n)$.}\BibitemShut {Stop}%
\end{thebibliography}


\end{document}